# Density Functional Theory Calculations of the thermochemistry of the dehydration of 2-propanol

Eugene Stephane Mananga[1,2,3], Aissata Diop[1,4], Paulin Dongomale[1,5], Fambougouri Diane[1,6], and Hubertus van Dam[1*]

[1]Computational Science Initiative, Brookhaven National Laboratory, Upton, New York 11973, US

[2]PH. D Programs of Physics & Chemistry, Graduate Center, The City University of New York, New York 10016, US

[3]Department of Engineering, Physics, and Technology, Bronx Community College, The City University of New York, 2155 University Avenue, CPH 118, Bronx, New York 10453, US

[4]Community College of Philadelphia, 1700 Spring Garden Street, Philadelphia, PA 19130, US

[5]Gore Nitrogen Pumping Service, 916 North Elm, Seiling, OK 73663, US

[6]Tufts University, 419 Boston Ave, Medford, MA 02155, US

**Abstract:**

Electronic structure theory provides a foundation for understanding chemical transformations and processes in complex chemical environments1[1]. Our work is focused on the NWChemEx project that has selected two interrelated science challenges that address the production of advanced biomass-derived fuels and other value-added chemical compounds[2]. One of which is the dehydration of 2-propanol over a zeolite catalyst. Aqueous phase dehydration of 2-propanol was investigated using density functional theory (DFT) calculations. We considered and analyzed the thermochemistry of the dehydration of 2-propanol using NWChem calculations while the NWChemEx code is still under development. Realistically modeling the reaction in this study properly requires simulations using extended atomistic models. We validated our computational models by comparing the predicted outcomes for 2-propanol dehydration with the calculated results from 1-propanol dehydration studies.[3] We used the first-principles DFT calculations to investigate aqueous phase dehydration of 2-propanol, examine the enthalpy of the 2-propanol reaction and computed the energy for geometry optimization for increasingly better basis sets: cc-pVDZ, cc-pVTZ, cc-pVQZ, cc-pV5Z, and cc-pV6Z[4]. The various transition states and minima along the reaction pathway are critical to inform the NWChemEx science challenge calculations. In this work, we established how the accuracy of the calculations depends on the basis sets, and we determined what basis sets are needed to achieve sufficient accurate results. We also calculated the reaction free energy as a function of temperature as thermodynamic parameter. We found that at low temperature the reaction is thermodynamically unfavorable. Nevertheless, dehydrating 2-propanol increases entropy, underscoring the need for high temperatures to facilitate the reaction. We have also started to investigate the reaction mechanism at the active site of the Zeolite H-ZSM-5 and how detailed electronic structure calculations can be used to develop a deeper understanding of catalysts, which might be improved. NWChemEx integrates a comprehensive suite of modules tailored for a wide range of electronic structure theory calculations prevalent in current research.[2]

We are working on setting up measures for chemical systems embedded in a larger system. This study explores how DFT can elucidate the dehydration of organic compounds like 2-propanol into new chemicals using zeolite catalysts. Our goals are to lay the groundwork for the mechanistic simulations and study the energy at different stages and transition states. Further insights into catalytic properties could be obtained in the future by comparing DFT calculations to other corresponding methods of predictions. Our results show how the enthalpy of each basis set decreases as the number of valence orbitals increases. These findings will form the basis for comparing calculated structures with experimental data. This work will explore and inform the NWChemEx development roadmap. It will also inform the ab-initio molecular dynamics simulations that will be used to validate the methods that NWChemEx implements.

Corresponding author: Dr. Hubertus van Dam, Email: hvandam@bnl.gov

## 1. Introduction

Contemporary computational chemistry merges principles from computer science, mathematics, physics, chemistry, materials science, and occasionally biology [5-8]. It is at the interface between these disciplines where many of the most exciting new developments occur. Nowadays, scientific problems require more from theories, computational algorithms and scientist's intuition than in previous years. Typically, experimental setups can accommodate larger systems than those we can currently simulate computationally. However, with computational methods we have more control over the chemical system under investigation. Also, we have direct access to the structure, so we can easily correlate structure and signals observed in an experiment. Recently, DFT has emerged as a powerful tool, surpassing experimental capabilities in determining structures and predicting material properties at unprecedented resolutions and scales. DFT is currently the most widely applied method in electronic structure theory[4]. The integrated NMR/DFT approach presents exciting advancements for clarifying the chemical bonds in materials, including ZSM-5 zeolites, the focal material of this study[9-13].

Our work focuses on the NWChemEx project, which aims to develop an exascale computational chemistry code inspired by two pivotal scientific challenges. These two science challenges are related to protein function and zeolite catalysis. As part of the project, an assessment of the computational performance as well as an accuracy of the calculations is required[4]. In this project, we are evaluating the characteristics for the zeolite science challenge. We also sought to assess how well the code performs across various hardware architectures. In addition, progress permitting, we will also explore aspects of the catalytic processes occurring in the zeolite. By investigating the zeolite, we also seek to derive universal fundamental insights into the structure and composition in superionic materials. Catalysts form the bedrock of the chemical industry, with their products integral to virtually all aspects of human activity[9]. At the core of catalysts lies the intricate relationship between their atoms and electrons, where quantum mechanics dictate interactions with reactants, products, and electromagnetic fields. Our work consists of using DFT to understand the mechanisms by which hydrogen is removed from organic compounds to form a new chemical. Our goals are to lay the groundwork for the mechanistic simulations and study the energy at different stages as well as different transition states. Further insights into catalytic properties could be obtained by comparing DFT calculations to other corresponding methods of predictions. The investigations performed in this project are examples of how detailed electronic structure calculations can be used to develop a deeper understanding of catalysts. We used DFT to perform initial calculations on the investigation of the dehydration of 2-propanol to understand the mechanisms by which water is removed from organic compounds to form a new compound over H-ZSM-5 zeolite. We obtained the data of the dehydration of 2-propanol, propene, water, and enthalpy using five (5) correlation-consistent basis sets. We simulated the aqueous phase dehydration of 1-propanol over H-ZSM-5 zeolite[3]. The results showed how the enthalpy of each basis set decreases as the number of valence orbitals increases. We are working to set up calculations for chemical systems that are embedded in a larger system. The zeolite case that we are interested in is an example. The catalysis ZSM-5 zeolite is used in the conversion of biomass into other products.

Today's challenge in organic compound structural determination requires a multifaceted approach that interface with many subjects such as physics, mathematics, computational algorithm, materials science and in some cases biology when needed. Each technique on its own has some limitations. However, with a new promising technique that seems to combine more than one phenomenon has become a common tool to unravel or elucidate the structure of organic compounds, such as proteins, carbohydrate, etc. DFT is such a computational method under consideration in my proposed work. This method will be used to investigate and elucidate the chemical bonds and structure of a catalyst in this project.

## 1.2. NWChem

A quantum chemistry code called NWChem was developed in 1994 and designed for parallel computers with the capability of implementing many energy expressions including DFT, and Hartree-Fock. The NWChem contains an umbrella of modules that can be used to tackle most electronic structure theory calculations being carried out today[14]. The following Fig. 3-01 is the modular design of NWChem.

- At the top we have "General Tasks"
- Then we have "Molecular Calculation Modules"
- These sit on the top "General Objects"
  - Geometry
  - Basis set
  - Etc.
- At the bottom is the parallelization infrastructure

**Figure 1-01. Modular design of NWChem.**

NWChem is an ab initio computational chemistry software package which includes quantum chemical and molecular dynamics functionality. However, the NWChem code was designed for computer architectures that no longer exist. NWChemEx is a complete rewrite of the software, which is modularized and flexible from its initial design and uses the power of C++. The primary focus of NWChemEx consists in designing the code that allows to run on graphic processing units (GPUs) and uses dynamic methods to shift computational work to optimize performance. The key points are modernization of the DFT code that is widely used in quantum chemistry, and the Coupled Cluster (CC) theory that is used to access the high accuracy required for predictive chemistry. The NWChemEx project is a natural extension of NWChem to overcome the scalability challenges associated with the migration of the current code base to exa-scale platforms. On a recent review by Apra et al. including the last author of this article[15], as above mentioned, it is highlighted that NWChemEx is being developed to address two outstanding problems in advanced biofuels research: (i) development of a molecular understanding of proton-controlled membrane transport processes and (ii) development of catalysts for the efficient conversion of biomass-derived intermediates into biofuels, hydrogen, and other bioproducts. Therefore, the main focus is on enabling scalable implementations of the ground state canonical CC formalisms utilizing the Cholesky decomposed form of the two-electron integrals, as well as linear scaling CC formulations based on the domain-based local pair natural orbital CC formulations and embedding methods.

NWChem provides a wide range of capabilities that can be deployed on supercomputing platforms to solve two fundamental equations of quantum mechanics, time-independent and time-dependent Schrödinger equations,

$$H|\Psi\rangle = E|\Psi\rangle, \qquad (1\text{-}1)$$

$$i\hbar \frac{\partial |\Psi\rangle}{\partial t} = H|\Psi\rangle, \qquad (1\text{-}2)$$

and a fundamental equation of Newtonian mechanics,

$$F_i = m_i a_i, \qquad (1\text{-}3)$$

where forces Fi include information about quantum effects[14]. Equations (3-1) and (3-2) can be addressed through different quantum mechanical representations, such as wavefunctions ($|\Psi\rangle$), electron densities ($\rho(r)$), and self-energies ($\Sigma(\omega)$). These representations form the core of NWChem's capabilities, enabling the computation of electronic wavefunctions, densities, and related properties for both molecular and periodic systems. These functionalities include Hartree–Fock self-consistent field (SCF)[14]. NWChem offers comprehensive DFT capabilities, incorporating both Gaussian and plane wave basis sets. Within the Gaussian basis set framework, a broad range of DFT response properties, ground and excited-state molecular dynamics (MD), linear-response (LR), and real-time (RT) time-dependent density functional theory (TDDFT) are available. The plane wave DFT framework enables scalable ab initio and molecular dynamics (MD) simulations. The plane wave code supports both norm-conserving and projector augmented wave (PAW) pseudopotentials. NWChem contains energy-gradient implementations of most exchange correlation functionals available in the literature, including a flexible framework to combine different functionals. The DFT module reuses elements of the Gaussian basis SCF module for the evaluation of the Hartree–Fock exchange and the Coulomb matrices by using 4-index 2-electron electron repulsion integrals. The formal scaling of the DFT computation can be reduced by choosing to use auxiliary Gaussian basis sets to fit the charge density and use 3-index 2-electron integrals instead. The DFT module accommodates both distributed data and mirrored array methods for assessing exchange-correlation potential and energy[14].

### 1.2.1. Density Functional Theory Calculations

NWChem uses Kohn-Sham[15] method for DFT, the implementation of that is very similar to the Hartree-Fock method, except that the exchange term is replaced by the exchange-correlation functional,

$$E_{xc} = \int f(\rho(r)) dr, \quad (1\text{-}4)$$

where $\rho(r)$ is the electron density and the simplest functional is,

$$f(\rho(r)) = [\rho(r)]^{4/3} \quad (1\text{-}5)$$

There are many functionals of increasing complexity[16], and we are using the DFT with the PBE (Perdew, Burke, Ernzerhof) functional[17,18]. NWChem implements numerous exchange and correlation terms, allowing for versatile combinations. The DFT integration is done numerically and need a grid of points and weights $r_i, w_i$, with the exchange correlation energy calculated as,

$$E_{xc} \approx \sum_i w_i f(\rho(r_i)) \quad (1\text{-}6)$$

### 1.2.2. COSMO Solvation Model and Implementation of Neural Network Potentials

While DFT calculations can provide highly accurate results the calculations are also relatively expensive. The recent development of machine learning methods has led to the development of neural network potentials (NNPs)[19] to predict molecular total energies and forces on the atoms from just the atomic positions. In practice the application of NNPs is orders of magnitude less compute intensive than the DFT calculations they replace, while providing results of a similar accuracy. Important for the development of NNPs is that they have to be designed to preserve the translational, rotational, and permutational symmetries of the system[20]. This is a stringent requirement for an NNP to correctly represent the Potential Energy Surface (PES)[21]. In our work we have used the recently developed DeePMD approach[22]. The following Fig. 3-02 is our NNP model requirements.

| NNP Model Requirements | | | | | |
|---|---|---|---|---|---|
| Molecular System | Data Collection | | Train | Test | Do it again! |
| Chosen Machine Learning method: DeePMD and NNP to simulate a number of typical chemical systems. | Prepare dataset for our molecule 2-propanol close to quantum chemistry method within NWChem. | input element information, atomic coordinate, charges, and spin state. | Train with TensorFlow framework. | Test set and training of neural network. PES for molecular system is predicted by neural network. | Use NNP on new molecular systems. |

**Figure 1-02. NNP model requirements.**

The dehydration of alcohols is an essential class of oxygen-removal reaction involved in the catalytic transformation of biomass-derived alcohols. During the reaction, as ions are stabilized due to the presence of water, for sensible predictions we need to account for the effects of water but, adding water molecules explicitly leads to very expensive calculations hence the purpose of using the COSMO (COnductor-like Screening Model)[23] method within NWChem to model and calculate 2-propanol with water. COSMO is not only cheaper but provides a better approximate and quicker calculation of 2-propanol in water.

As we strive for a healthier earth with carbon-neutral alternatives, removal of oxygen and alcohol (-OH) functional groups is essential to the transition of biomass into fuels. We used NWChem to minimize our large molecular model, the HZSM-5 zeolite. A piece around the active site (Al-OH) was carved out, the fragment was terminated then optimized to add 2-propanol. As we advance in our investigation, we are replacing COSMO and DFT calculations for a more efficient and cheaper calculation of 2-propanol and other molecules. COSMO calculations such as DFT calculations are slow as they require two steps, one DFT calculation of the molecular system in vacuum, and another DFT calculation starting from the vacuum solution and incorporating the solvation effects[23]. As an alternative, we are replacing DFT calculations with our chosen machine learning method[24], DeePMD[22], and neural network potentials (NNPs)[19] to stimulate several chemical molecular systems. A neural network model is 100,000 cheaper and facilitates the calculation of transition states and can replace SCF gradient calculations with NNPs to predict forces and positions of atoms. Our goal is to use this NNP on new molecular systems again to stabilize various compounds in water. We replicated the training model for methane using DeepMD and Python which resulted in favorable data about the energy and force. As a result, NNP will facilitate the calculations of other molecules to stabilize various compounds in water and preserve the translational, rotational, and permutational symmetries of the system. We are preparing a dataset for 2-propanol close to NWChem, which includes information of the elements, atomic coordinate, charges and spin state to test and train with Tensor Flow framework then test and train the neural network Potential Energy Surface for molecular systems predicted by the neural network.

### 1.3. Catalysts

A catalyst is an intermediary that assists in the conversion of many raw materials to a particular usable form and commonly used in industry[9]. This project investigates the ZSM-5 zeolite catalyst, which is the trade name for zeolite Socony mobil-5, a pentasil aluminosilicate zeolite. Each pentasil unit is made up of eight 5-membered rings, and individual pentasil units are connected to each other by oxygen bridges to form regular corrugate sheets with 10-ring holes. The channel structure and acidity of ZSM-5 lends itself to several industrial uses and a support material for catalysis. ZSM-5 is a synthetic zeolite which contains silica (Si) and alumina (Al) with the ratio of silica greater than the alumina. The catalyst is known as ZSM-5 because it has a pore diameter of 5 angstroms, and it has more than five (5) Si/Al ratios. Catalyst ZSM-5 zeolite is used in the conversion of 2-propanol, by removal of water molecules[25]. In this investigation, DFT provides an insight into the mechanism by which the water is removed when the catalyst comes into contact with 2-propanol alcohol. We are working to set up calculations for chemical systems that are embedded in a larger system. The catalytic ZSM-5 zeolite is used in the conversion of biomass into other products. Catalysts are the cornerstone of the chemicals industry, whose products are used in nearly all human endeavors[9]. We ran some calculations to see how well the NMR capabilities of NWChem would work on a large system like the zeolite cluster. Using 16 cores the calculation on this system with 115 atoms finished in under 3 hours using the computationally efficient 3-21g* basis set[26]. NMR might be a way to monitor the chemical reactions experimentally. Solid state NMR capabilities would have to be developed within NWChem to be able to simulate such experiments for extended systems such as zeolites. Currently NWChem has only got NMR capabilities for finite systems be they molecules or fragments of extended systems.

### 1.4. Mechanistic Simulations of Zeolite H-ZSM-5

Electronic structure theory provides a foundation for understanding chemical transformations and processes in complex chemical environments. This research is focused on the NWChemEx that has selected two interrelated science challenges that address the production of advanced biomass-derived fuels and other value-added chemical compounds. One of which is the dehydration of 2-propanol over a zeolite catalyst. Aqueous phase dehydration of 2-propanol was investigated using density DFT calculations. We considered and analyzed the thermochemistry of the dehydration of 2-propanol using the NWChemEx calculations.

Modeling computational challenges of the reaction in this study properly requires simulations using extended atomistic models. We validated the computational outcomes of 2-propanol by comparing the predictions against the dehydration of 1-propanol. We used the first-principles DFT calculations to investigate aqueous phase dehydration of 2-propanol, examine the enthalpy of the 2-propanol reaction and computed the energy for geometry optimization for increasingly better basis sets: cc-pVDZ, cc-pVTZ, cc-pVQZ, cc-pV5Z, & cc-pV6Z. Identifying the transition states and minima along the reaction pathway is crucial for informing the calculations related to the NWChemEx science challenges.

### 1.4.1. E1 and E2 Mechanisms

The authors Lepore et al.[27], Mei et al.[3], or Zhi et al.[28] have discussed the intramolecular dehydration of 1-propanol which proceeds via an E1 or E2 type mechanism.

**Figure 1-03. E2 Mechanism of Dehydration of 1-propanol.**

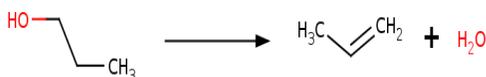

**Figure 1-04. First step in E1 Mechanism of Dehydration of 1-propanol.**  **Figure 1-05. Second step in E1 Mechanism of Dehydration of 1-propanol**

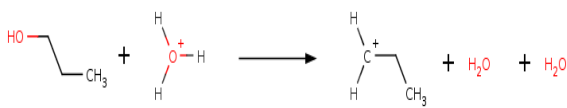 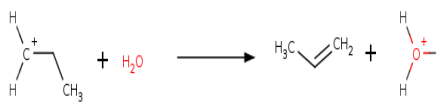

Other authors such as Car et al. have proposed the intermolecular SN1 or SN2 type mechanism for alcohol dehydration[29]. Let us recall that there are two types of elimination reactions, E1 and E2, which consist of different number of steps[30]. The most obvious way to distinguish E1 versus E2 is by looking at the number of steps in the mechanism. The E1 elimination reaction has a two-step mechanism that involves a carbocation intermediate. E1 elimination favors the most substituted alkene product. On the other hand, E2 takes place in one step and has no intermediate. This entire process takes place in one step, unlike the E1 reaction, which takes place in two steps. An elimination reaction consists of a formation of a C-C double bond by the removal of 2 single bonds from adjacent carbons; a leaving group attached to a carbon and a hydrogen attached to an adjacent carbon will usually be the two bonds broken in these types of reactions. In the E1 mechanism, the protonated oxygen leaves as a water molecule to generate a cationic intermediate that subsequently undergoes deprotonation to form the alkene. In the E2 mechanism, loss of water and deprotonation occur simultaneously, and no intermediates are involved.

Understanding the mechanism of dehydration is important to build models for organic reaction pathways under some relevant thermodynamics' conditions. Here, we described a detailed study of the mechanism of E1 is associated with a negative entropy of activation, whereas the E2 mechanism is associated with a positive entropy of activation. This suggests that E2 elimination may become less favorable compared to other possible competing reactions at higher temperatures, whereas the E1 mechanism may become more competitive. The lack of reagents and the use of water as the solvent mean that this reaction conforms to many of the standard principles of green chemistry. Bockisch et al. highlighted that hydrothermal dehydration of alcohols is also of interest as an alternative to conventional dehydration of alcohols at ambient conditions[30]. Alcohol dehydrations performed at ambient conditions require concentrated acids and the use of forcing conditions, such as physical separation of reactants and products via distillation. Hydrothermal reactions of organic chemicals in general are of increasing interest for green chemistry applications[30]. In the present work, the dehydration mechanisms of 2-propanol (over H-ZSM-5 zeolites) at various temperatures in both the absence and presence of various water concentrations was studied using DFT calculations. In particular, the effects of the formation of the propene and water on the reaction pathways were investigated. Our calculated results obtained were compared to the experimentally observed and computed results of 1-propanol in recent works[3].

### 1.4.2. Methods

#### 1.4.2.1. DFT with the PBE functional (PBE is the Perdew, Burke, Ernzerhof functional)

We performed molecular DFT calculations within the generalized gradient approximation (GGA) with the exchange correlation functional of Perdew, Burke, and Ernzerhoff (PBE)[17] as implemented in the NWChem package[14]. GGA for the exchange-correlation energy was shown to improve upon the local spin density (LSD) description of atoms, molecules, and solids. Test calculations show that the error due to the basis is less than 2.62 kcal/mol which is acceptable. Chemical accuracy is an error of less than 1 kcal/mol but then this applies to the reaction enthalpy. As we are making errors of the same sign due to the basis set incompleteness in both reactants and products, the error in the reaction enthalpy overall is smaller than the error we are making in either reactants or products. Due to the variational principle that applies to DFT calculations, the energy we get is always too high due to basis incompleteness. In other words, when we calculate the reaction enthalpy as a difference of the energies of the molecules at either side of the reaction arrow, the basis set incompleteness error will cancel to some degree. In the next section, we show the results of the calculation of the energies of the three molecules in the reaction. We considered how these energies depend on the basis sets and temperatures. We used a family of basis sets where one basis set extends another in a systematic way. Hence the further you go to the right the better the basis set becomes in all respects. Because the Hartree-Fock and DFT are both variational methods, we achieve the lowest energy for the best basis sets. This information is used to look at how the energy converges as a function of the basis set. Note that chemistry is mainly about energy differences. In this case, the errors introduced by the basis set might cancel when the difference is computed. In other words, the reaction enthalpy might converge faster as a function of the basis set than the energies of the individual molecules. Initially, for 2-propanol, we obtained a reaction energy change of + 0.02672780 Hartree. This data didn't account for the thermodynamic effects, so it is possible that including the thermo contributions the reaction goes from endothermic to exothermic due to the entropy. We did a test with the calculations in vacuum where the reaction is slightly endothermic (i.e., it costs energy). However, we didn't account for the presence of water. It is possible that a water molecule binds stronger to water than 2-propanol does. In addition, initially, we didn't account for the entropy effects either. Molecular vibrational modes can provide estimates of entropy effects. This is the next aspect we look at. Regarding the cc-pV6Z basis set calculations with the NWChem, some linear dependencies for both 2-propanol and propene is detected. For example, in the 2-propanol output there is the statement "WARNING: Found 11 linear dependencies." While NWChem has methods to limit the impact of these linear dependencies they are not fool proof, and it is best to avoid these kinds of cases. Next, we compared our calculations of 2-propanol with the results and plots of 1-propanol by Mei and Lercher[3]. Mei and Lercher' diagrams show that for 1-propanol the reaction is exothermic. i.e. forming propene and water lowers the energy by 188 kJ/mol. NWChem outputs energies in the unit Hartree, 1 Hartree is 2625.5 kJ/mol, so lowering the energy by 188 kJ/mol is equivalent to an energy change of -0.071605 Hartree.

### 1.4.2.2. Geometry optimizations

We have performed a geometry optimization with increasingly better basis sets of "correlation-consistent"[31]: cc-pVDZ, cc-pVTZ, cc-pVQZ and cc-pV5Z. and cc-pV6Z. This family of cc basis sets is systematic in the sense that going from cc-pVDZ to cc-pV6Z, every step to the next basis set takes the previous basis set and adds more functions. In computational chemistry, a basis set is a set of functions that is used to represent the electronic wave function in the Hartree-Fock method or DFT to turn the partial differential equations of the model into algebraic equations suitable for efficient implementation on a computer. Note that cc basis sets are built up by adding shells of functions to a core set of atomic Hartree-Fock functions. Each function in a shell contributes very similar amounts of correlation energy in atomic calculation. For the $1^{st}$ and $2^{nd}$ row atoms, the cc-pVDZ basis set adds 1s, 1p, and 1d function. The cc-pVTZ set adds another s, p, d, and an f function, etc… Various augmentations to these base sets have been developed such as the addition of diffuse functions to better describe anions and weakly interacting molecules (aug-cc-pVnZ) as well as special basis sets designed for describing the effects of correlating the core electrons. This set of basis sets and approximations was chosen for convenience because it spans the space of complexity in common density functionals. These geometry optimizations fix some atoms in place to model a fixed environment that constrains how the atoms can move during the optimization of an active part of a chemical system.

### 1.4.2.3. Transition state searches

The transition states correspond to saddle points on the potential energy surface, which is a first order stationary point along the reaction coordinate. The first order saddle points are maxima in one direction (along the reaction coordinate) and minima in all other directions. The transition state search attempts to localize stationary points with one negative second derivative, identifying the reaction mode, and maximizing energy in all directions. The optimization of the transition state is more challenging than the search for stable structures, and search for a transition state should start near the transition state.

### 1.4.3. Thermochemistry calculations

We generated the thermochemical information from Ochterski[32,33] who ran calculations at 1.0 atmosphere and 298.15 K for each of the reactants and products in the reaction where ethyl radical abstracts a hydrogen atom from molecular hydrogen:

$$C_2H_5 + H_2 = C_2H_6 + H \qquad (1\text{-}7)$$

as well as for the transition state[34]. The thermochemistry output from Gaussian is summarized in Table 1.

|  | $C_2H_5$ | $H_2$ | $C_2H_6$ | H | C |
|---|---|---|---|---|---|
| $\varepsilon_0$ | -77.662998 | -1.117506 | -78.306179 | -0.466582 | -37.198393 |
| $\varepsilon_{ZPE}$ | 0.070833 | 0.012487 | 0.089704 | 0.000000 | 0.000000 |
| $E_{tot}$ | 0.074497 | 0.014847 | 0.093060 | 0.001416 | 0.001416 |
| $H_{corr}$ | 0.075441 | 0.015792 | 0.094005 | 0.002360 | 0.002360 |
| $G_{corr}$ | 0.046513 | 0.001079 | 0.068316 | -0.010654 | -0.014545 |
| $\varepsilon_0 + \varepsilon_{ZPE}$ | -77.592165 | -1.105019 | -78.216475 | -0.466582 | -37.198393 |
| $\varepsilon_0 + E_{tot}$ | -77.588501 | -1.102658 | -78.213119 | -0.465166 | -37.196976 |
| $\varepsilon_0 + H_{corr}$ | -77.587557 | -1.101714 | -78.212174 | -0.464221 | -37.196032 |
| $\varepsilon_0 + G_{corr}$ | -77.616485 | -1.116427 | -78.237863 | -0.477236 | -37.212938 |

Table 1: Calculated thermochemistry values from Gaussian for the reaction $C_2H_5 + H_2 \rightarrow C_2H_6 + H$. All values are in Hartrees.

Using the calculation of $H_2$ from Hartree-Fock with the STO-3G basis set, we have generated numbers that are very close to those obtained by Ochterski got from Gaussian calculation[33]. These quantities include: $\varepsilon_0$, $\varepsilon_{ZPE}$, $E_{tot}$, and $H_{corr}$. In the following, we have summarized the key lines from the output file:

```
From line 2735 and a little bit further I have:

            Total SCF energy =     -1.117505881567
          One-electron energy =    -2.540621409644
          Two-electron energy =     0.680067418611
       Nuclear repulsion energy =   0.743048109466

And from the lines 3022 and further:

    Temperature                     =    298.15K
    frequency scaling parameter     =    1.0000

    Linear Molecule

    Zero-Point correction to Energy  =    7.833 kcal/mol  (   0.012483 au)
    Thermal correction to Energy     =    9.314 kcal/mol  (   0.014843 au)
    Thermal correction to Enthalpy   =    9.906 kcal/mol  (   0.015787 au)

    Total Entropy                    =   30.951 cal/mol-K
       - Translational               =   28.068 cal/mol-K (mol. weight =   2.0156)
       - Rotational                  =    2.884 cal/mol-K (symmetry #  =        2)
       - Vibrational                 =    0.000 cal/mol-K

    Cv (constant volume heat capacity) =  4.966 cal/mol-K
       - Translational               =    2.979 cal/mol-K
       - Rotational                  =    1.986 cal/mol-K
       - Vibrational                 =    0.000 cal/mol-K
```

From these results we can match the quantities Ochterski used with the quantities provided in the NWChem output. We find that $\varepsilon_0$ is the total electronic Energy, $\varepsilon_{ZPE}$ is the Zero-Point correction to Energy, $E_{tot}$ is the correction to the internal thermal energy ($E_{tot} = E_{trans} + E_{rot} + E_{vib} +$ $E_{elec}$) and H$_{corr}$ is the correction to the thermal Enthalpy. The entropy contribution can be calculated using the relation[33]:

$$S = Nk_B \left[1 + \ln\left(\frac{q(V,T)}{N}\right) + T\left(\frac{\partial (\ln q)}{\partial T}\right)_V\right] \qquad (1\text{-}8)$$

where q (V, T) is the partition function for the corresponding component of the total partition function, V the volume and T is temperature. Molar values are given by, $n = \frac{N}{N_A}$, and the gas constant, $R = N_A k_B$, where $N_A$ and $k_B$ are the Avogadro number and the Boltzmann constant, respectively. The total entropy is: $S_{tot} = S_{trans} + S_{rot} + S_{vib} + S_{elec}$. Gaussian assumes that the first electronic excitation energy is much greater than $k_B T$. Therefore, the first and higher excited states are assumed to be inaccessible at any temperature. Furthermore, the energy of the ground state is set to zero. These assumptions simplify the electronic partition function. Since there are no temperature dependent terms in the partition function, the electronic heat capacity and the internal thermal energy due to electronic motion are both zero. The change of the entropy contribution is approximated to:

$$\Delta S_{tot} = \Delta S_{trans} + \Delta S_{rot} + \Delta S_{vib} + \Delta S_{elec} \approx \Delta S_{trans} + \Delta S_{rot} + \Delta S_{vib} \qquad (1\text{-}9)$$

where $\Delta S_{trans}, \Delta S_{rot}, \Delta S_{vib}$, and $\Delta S_{elect}$ are the translational, rotational, vibrational, and electronic entropy contributions to the total entropy $\Delta S_{tot}$, respectively. The internal thermal energy (E$_{tot}$) can also be written in terms of the partition function (Ref. McQuarrie)[33]:

$$E_{tot} = Nk_B T^2 \left(\frac{\partial \ln q}{\partial T}\right)_V \qquad (1\text{-}10)$$

These above equations are used to derive the final expressions used to calculate the different components of the thermodynamic quantities obtained by Gaussian. Note that the equations used for computing thermochemical data in Gaussian are equivalent to those given in standard texts on thermodynamics[33]. The contribution to the internal thermal energy due to translation is:

$$E_{trans} = \frac{3}{2} k_B T \qquad (1\text{-}11)$$

The contribution to the internal thermal energy due to rotation is:

$$E_{rot} = \frac{3}{2} k_B T \qquad (1\text{-}12)$$

The contribution to the internal thermal energy due to vibration is:

$$E_{vib} = \sum_i \frac{h\gamma_i}{\left(e^{h\gamma_i/k_B T} - 1\right)} \qquad (1\text{-}13)$$

The zero-point energy (ZPE) is:

$$ZPE = \sum_i \frac{h\gamma_i}{2} \qquad (1\text{-}14)$$

where h and $\gamma_i$ are Planck constants and the vibrational frequency. The total entropy contribution terms from the translational, rotational, vibrational, and electronic terms can be explicitly calculated and we are providing the results obtained by McQuarrie[33]:

$$S_{trans} = k_B \left\{\ln\left[\left(\frac{2\pi m k_B T}{h^2}\right)^{3/2} \frac{k_B T}{P}\right] + \frac{5}{2}\right\} \qquad (1\text{-}15)$$

$$S_{elect} = R \ln q_e \qquad (1\text{-}16)$$

$$S_{rot} = k_B \left\{\ln\left[\frac{\sqrt{\pi I_A I_B I_C}}{\sigma}\left(\frac{8\pi^2 k_B T}{h^2}\right)^{3/2}\right] + \frac{3}{2}\right\} \qquad (1\text{-}17)$$

$$S_{vib} = k_B \sum_i \left( \frac{h\gamma_i}{k_B T \left(e^{h\gamma_i/k_B T}-1\right)} - \ln\left(1 - e^{-h\gamma_i/k_B T}\right) \right) \qquad (1\text{-}18)$$

where P, m, $\sigma$, and $q_e$ are the pressure, mass, symmetry number, and the electronic partition function. $I_A$, $I_B$, and $I_C$ are the principal moments of inertia for the molecules of interest. To account for important entropic contribution ($\Delta S$) and zero-point energy (ZPE) corrections, both Gibbs free energy ($\Delta G$) and enthalpy ($\Delta H$) changes along reaction pathways were calculated using standard statistical thermodynamic method. In summary,

$$H = \varepsilon_0 + H_{corr}, \qquad (1\text{-}19)$$
$$\Delta H = H_{products} - H_{reactants}, \qquad (1\text{-}20)$$
$$\Delta G = \Delta H - T\Delta S = \Delta H - T(S_{products} - S_{reactants}). \qquad (1\text{-}21)$$

We carefully verify the units. The "Total DFT Energy" is given in Hartree. The correction to the thermal Enthalpy ($H_{corr}$) is also given in Hartree. The entropy is given in cal/mol.K, so these numbers need to be converted with a factor 1/(627.5*1000) to obtain the unit in Hartree/mol.K. These conversions allow to obtain the Gibbs free energy $\Delta G$ in the unit of Hartree/mol. Next, we need to multiply the result with 627.5 to get $\Delta G$ in kcal/mol. This approach is consistent with previously reported theoretical modeling work by McQuarrie and Simon[34], by Mei and Lercher[3] on the 1-propanol dehydration in the H-ZSM-5 zeolites, and by the work by Ochterski in the reaction where ethyl radical abstracts a hydrogen atom from molecular hydrogen as well as for the transition state[33].

### 1.4.4. Results and Discussion

Although the principal objective of this study has been to elucidate the convergence patterns of various standard correlation-consistent basis sets in the context of DFT applied to elimination reactions E1 and E2. We obtained data of the dehydration of 2-propanol, propene, water, and enthalpy using five correlation-consistent basis sets in three groups of temperatures as shown below. We have simulated the aqueous phase dehydration of 2-propanol without water and over water. The results showed how the enthalpy of each basis set decreases as the number of valence orbitals increases. We have investigated the mechanism of 2-propanol dehydration using variable temperatures as thermodynamic parameters. The calculated Gibbs free energy profile of 2-propanol dehydration via both E1 and E2 mechanisms in the unimolecular reaction route is shown in Figures 1-03 to 1-05. We presented a detailed analysis of the enthalpy and entropy contributions to the reactions. The E2 elimination mechanism dominates over the corresponding E1 mechanism, with the E2 mechanism being competitive with E1 in the present study. These results are relevant to understanding the kinetics and product distributions of alcohol dehydration reactions in natural chemical systems and can guide the development of organic chemical reactions that mimic organic reactions under laboratory green chemistry conditions. This is also consistent with the recently reported DFT calculations of 1-propanol in H-ZSM-5 and other types of zeolites.

### 1.4.5. Enthalpy Analysis

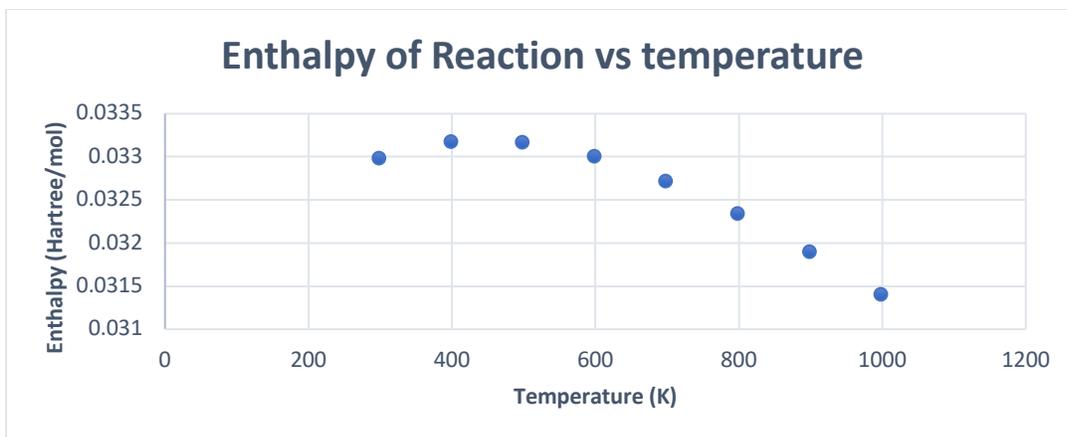

**Fig. 3-1.** Dehydration of 2-propanol for cc-pVDZ basis set with temperature in the range of 298.15K and 998.15 K.

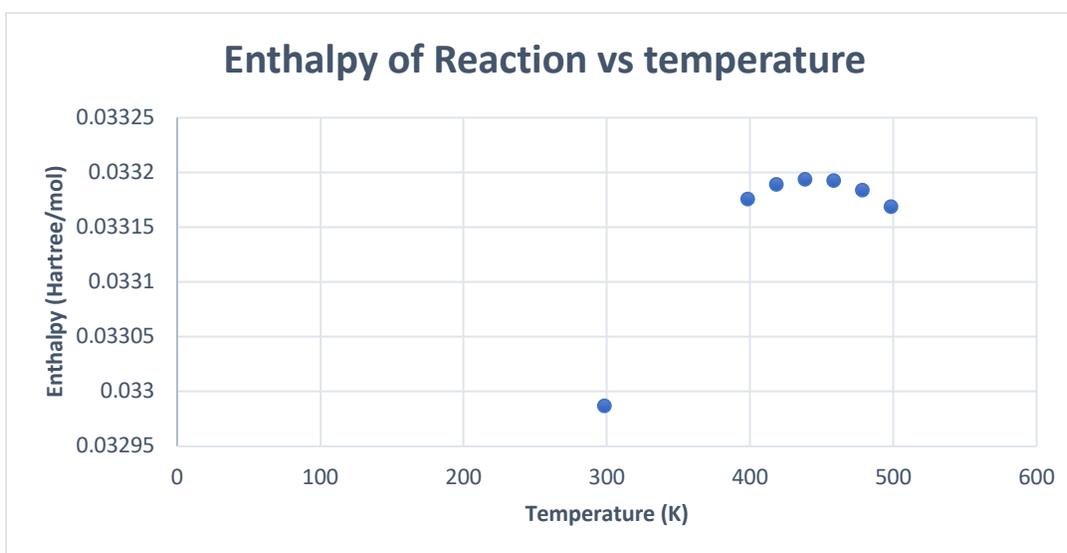

**Fig. 3-2.** Dehydration of 2-propanol for cc-pVDZ basis set with temperature in the range of 298.15K and 498.15 K.

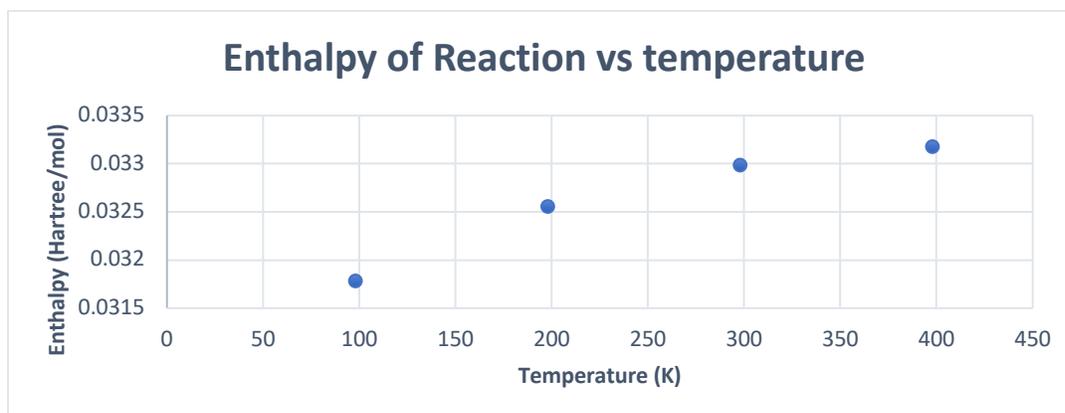

**Fig. 3-3.** Dehydration of 2-propanol for cc-pVDZ basis set with temperature in the range of 98.15K and 398.15 K.

The following graphs represent the enthalpy and enthalpy correction of the 1- and 2-propanol reaction in the range of temperature from 98.15 K to 998.15 K.

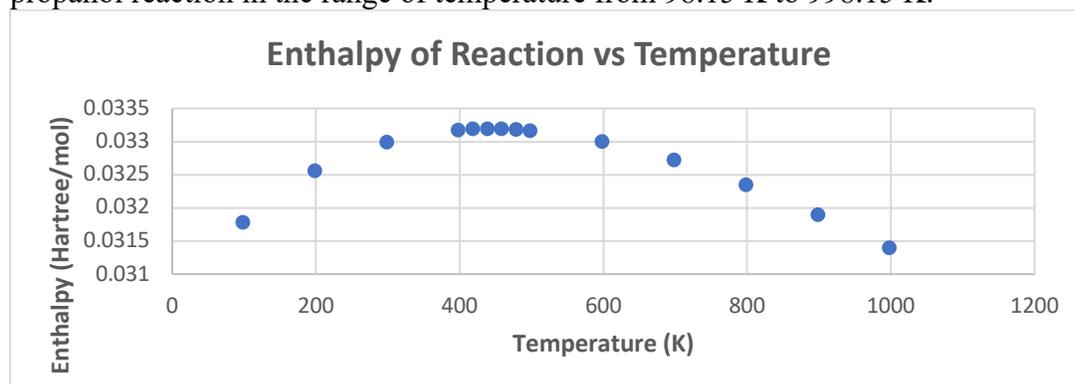

**Fig. 3-4. Dehydration of 2-propanol for cc-pVDZ basis set with temperature in the range of 98.15K and 998.15 K.**

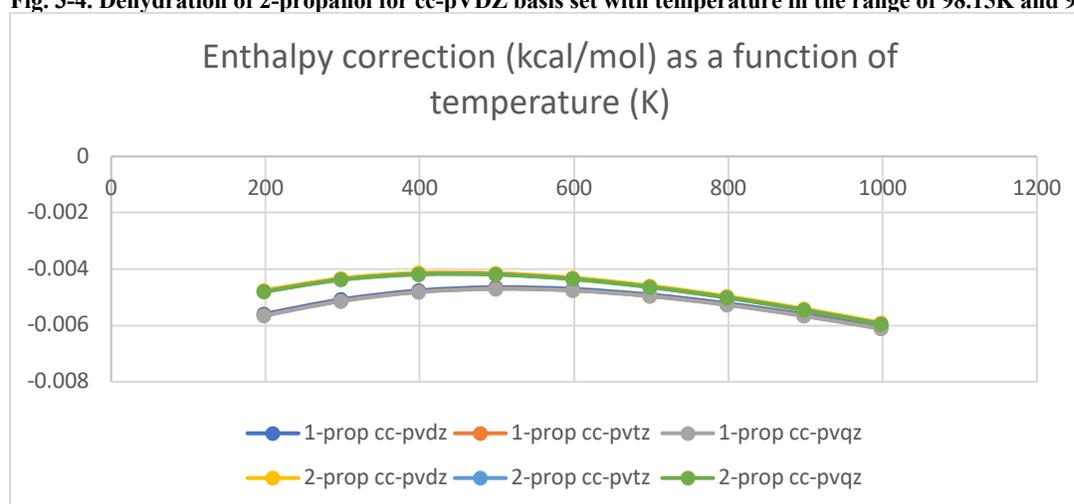

**Fig. 4. Combined dehydration of 1- and 2-propanol with temperature in the range of 198.15K and 998.15 K.**

**Enthalpy Analysis**

The enthalpy is a central factor in thermodynamics, and it is the heat content of a system. Looking at the dehydration of 2-propanol reaction at all temperatures that we have investigated, we hound a positive change of enthalpy, which means that the reaction is endothermic, where the energy is taken in from the surroundings. We have performed the thermochemistry calculations of 2-propanol for cc-pVDZ basis set in three groups of temperatures from 98.15 K to 998.15 K. In the first group of temperatures corresponding to Figure 3-1, the enthalpy increases from 0.032987 Hartree/mol (T = 298.15 K) to 0.033176 Hartree/mol (T = 398.15 K), then decreases to 0.031402 Hartree/mol (T = 998.15 K). A closer look at the range of temperature between 418.15 K and 498.15 K, with an increment of 20 K (corresponding to Figure 3-2) shows an increase of the enthalpy until a maximum of about 0.033194 Hartree/mol (T = 438.15 K), then a decrease to about 0.033169 Hartree/mol (T = 498.15 K). This range of temperatures from about 398.15 K to about 498.15 K is less favorable to the reaction since it requires more energy to produce the dehydration of 2-propanol reaction. In Figure 3-3, the enthalpy increases from 0.031781 Hartree/mol (T = 98.15 K) to 0.033176 Hartree/mol (T = 398.15 K). The increase of enthalpy in this range of temperature indicates that only the lowest temperature (around 98.15 K) is favorable for the 2-propanol reaction to happen. Overall, Figures 3-1 to 3-4 shows that the reaction is favored at both lower (around 98.15 K), and high temperature (around 998.15 K) because the enthalpy of the reaction is sufficiently decreased during these extreme temperatures. Figure 4 shows the enthalpy correction of the dehydration of 1- and 2- propanol versus the temperature in the range of 198.15 K to 998.15 K. The enthalpy correction is calculated as

$$H_{Corr} = E_{tot} + k_B T, \qquad (1\text{-}22)$$

where $E_{tot}$, $k_B$, and T are described above. The enthalpy correction as a function of temperature (Figure 4) is consistent with the enthalpy of reaction versus temperature (Figure 3). The enthalpy correction for 1-propanol is larger than 2-propanol for the same basis set and temperatures. The dehydration of 2-propanol requires less enthalpy correction than 1-propanol for the temperatures investigated below 1000K. This indicates once again that the 2-propanol reaction is less favorable to occur than 1-propanol reaction in the region of the temperatures we have investigated from 198.15 K to 998.15 K. Figure 4 shows that the calculations performed on the basis sets cc-pVDZ, cc-pVTZ, and cc-pvQZ present a small change of the enthalpy correction during the dehydration of 1- and 2- propanol.

### 1.4.6. Entropy Analysis

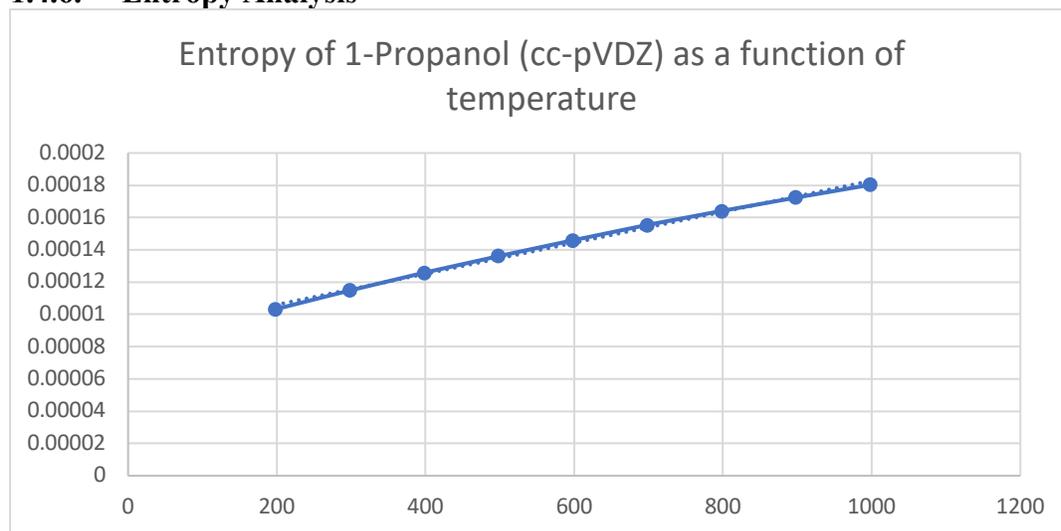

**Fig. 5. Dehydration of 1-propanol for cc-pVDZ basis set with temperature in the range of 198.15K and 998.15 K.**

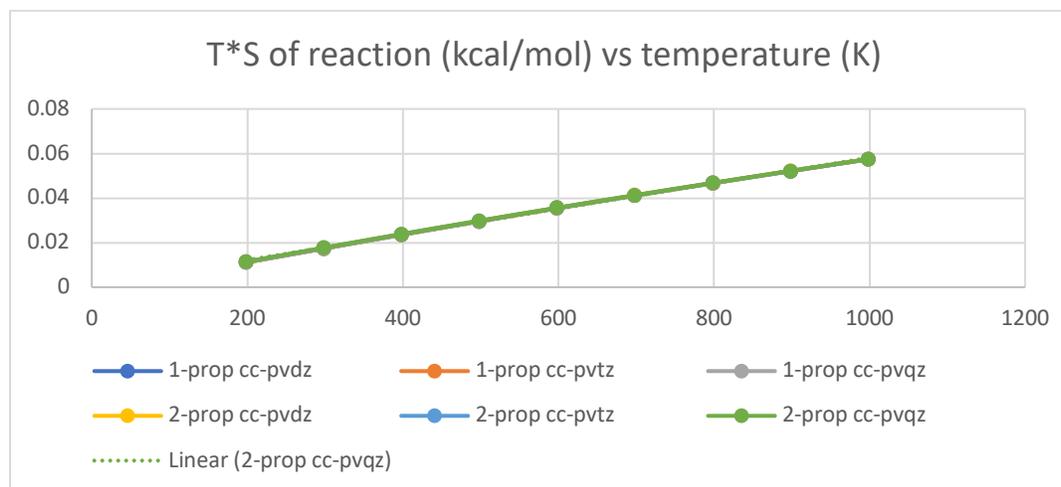

Fig. 6. Combined dehydration of 1- and 2-propanol with temperature in the range of 98.15K and 998.15 K.

**Entropy Analysis**

Unlike enthalpy, in physics, the entropy is described as a thermodynamic quantity representing the unavailability of a system's thermal energy for conversion into mechanism work, often interpreted as the degree of disorder in the system. Figures 5 and 6 show positive entropy, which means an increase in disorder during the dehydration of 1- and 2-propanol reactions. Both reactions 1- and 2-propanol occur with positive entropy or disorder. Furthermore, as the temperature increase, both reactions present nearly a linear increase of positive entropy (Figures 5, and 6), which means both reactions become more disordered making their change more spontaneous or favorable. However, a closer look at Figure 6 shows that at lower temperatures (about 200 K), the 2-propanol dehydration reaction is more disordered than the 1-propanol reaction. Also, at higher temperatures (about 1000 K), the same observation is made, while the 2-propanol reaction is slightly less disordered when performed in the range of temperatures from about 450 K to about 800 K. Because the reaction 2-propanol dehydration is more disordered at lower temperatures (about 200 K) and higher temperatures (about 1000 K), the reaction is more spontaneous and favorable at these extreme temperatures compared to the dehydration 1-propanol reaction.

### 1.4.7. Gibbs Free Energy Analysis

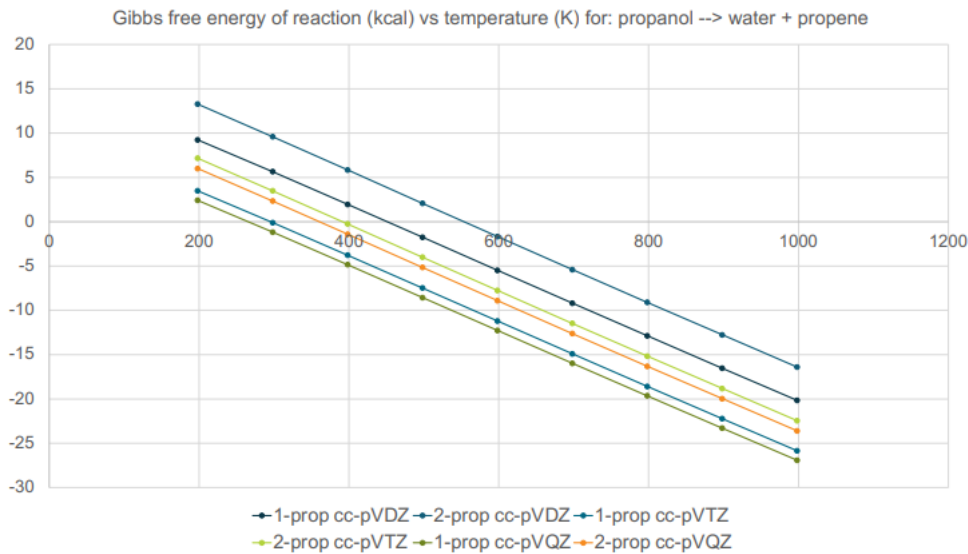

Fig. 7-1. Combined dehydration of 1-propanol and 2-propanol

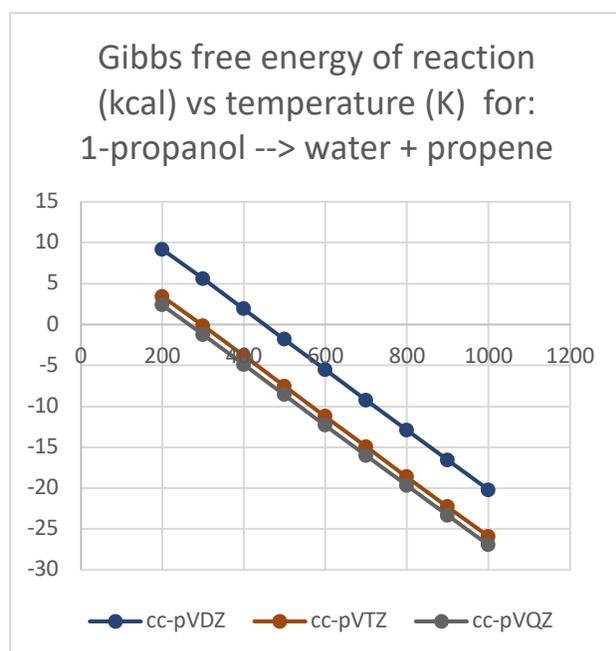
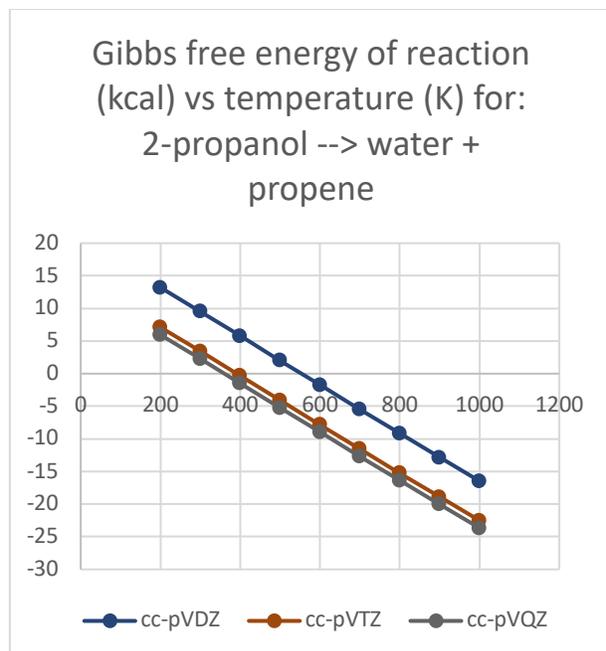

**Fig. 7-2.** Dehydration of 1-propanol.  **Fig. 7-3.** Dehydration of 2-propanol.

**Analysis and discussion**

    The Gibbs free energy is one of the most important thermodynamics functions for the characterization of a system. In isothermal, isobaric systems, the Gibbs free energy can be thought of as a "dynamic" quantity, which is a representative measure of the competing effects of the enthalpic and entropic driving forces involved in a thermodynamics process. As a necessary condition for the reaction to occur at constant temperature and pressure, $\Delta G$ must be smaller than the non-pressure-volume work, which is often equal to zero, i.e. $\Delta G$ must be negative. If the analysis indicates a positive $\Delta G$ for a reaction, then energy in the form of non-pV work would have to be added to the reacting system for $\Delta G$ to be smaller than the non-pV work and make it possible for the reaction to occur. The change of Gibbs free energy, $\Delta G$ can be considered as the amount of "free" or "useful" energy available to do non-pV work at constant temperature and pressure. The change $\Delta G$ in Gibbs free energy that is caused by the reaction is a quantitation measure of the favorability of a given reaction under a natural tendency to achieve a minimum of the Gibbs free energy.

    A reaction is permissible only if it results in a zero or positive total change in the universe's entropy. This is reflected in a negative $\Delta G$, and the reaction is an exergonic process. If the Gibbs free energy is negative, the reaction is random and will proceed forward and spontaneously. A negative $\Delta G$ means that the reactants, or initial state, have more free energy than the products or final state. This results in spontaneous reactions, which proceed without external energy input. Hence, the lower Gibbs free energy (i.e. more negative) means the equilibrium is pushed towards the product side. Naturally, a $\Delta G$ of zero indicates that the system is at equilibrium.

    Fig. 7-1 shows that the Gibbs free energy is very sensitive to the correlation-consistent basis sets, and to the change of the temperatures. In the region of temperature below 298 K (low

temperatures), the dehydration of 1- and 2- propanol are not favorable and spontaneous. Because the plots indicate the free energy $\Delta G$ of both reactions to be positive for low temperatures, the energy in the form of non-pV work would have to be added to 1- and 2- propanol for $\Delta G$ to be smaller than the non-pV work and make it possible for the reaction to occur. For the same cc basis set, positive $\Delta G$ is larger for 2-propanol than 1- propanol, which means that 1-propanol reaction would require less energy than 2-propanol to make the dehydration possible at low temperatures. Figs. 7-2 and 7-3 also show that each reaction, 1-propanol and 2- propanol, have a particular temperature, where $\Delta G = 0$, indicating that the system is at equilibrium.

As the temperature increases, we observed that both reactions become increasingly more favorable or spontaneous, which correspond to $\Delta G$ becoming more and more negative. The reactants 1- and 2- propanol have more free energy than water and propene, which lead to a favorable or spontaneous dehydration reaction without the addition of energy. Naturally, highest temperature (about 1000K) in the region of temperature we have investigated have the largest negative $\Delta G$, which correspond to more favorable or spontaneous dehydration reaction. All plots in Figs. 7-1, 7-2, and 7-3 remain linearly parallel in the region of temperatures that we have investigated the 1- and 2- propanol reactions. This means that the convenient correlation consistent basis sets (cc-pVDZ, cc-pVTZ, and cc-pVQZ) we have selected for our investigation act in a similar way during the DFT calculations. Furthermore, for the same temperatures applied to both reactions, the 1- propanol reaction remains more favorable or spontaneous than the 2- propanol in the regime of temperature we have investigated. This means that in the positive regime, $\Delta G$ is more positive for 2- propanol compared to 1- propanol and in the negative regime, $\Delta G$ is more negative for 1- propanol compared to 2- propanol.

Figs. 7-1, 7-2, and 7-3 represent the Gibbs free energy DFT results for E2 mechanism of dehydration of 1- and 2- propanol described by Figs 1-3 and 2-3. The E2 elimination reaction takes place in one step and has no intermediate. This is a second order bimolecular reaction, meaning that the rate of reaction depends on both the substrate or the concentration of the molecule, and the deprotonating or the assisting base. In the E2 reaction, the product is dependent on the stereochemistry of the molecule. The leaving group and hydrogen must be "anti" to each other, meaning they must rest on opposite sides of the molecule. Since the deprotonation and the ionization are happening at the same time in this reaction, for there to be enough space, the two parts of the molecule must be as far away from each other as possible. During the E2 dehydration mechanism of 1- and 2- propanol, the leaving proton ion H+ grabs onto the water molecule that is a part of the molecule. Then, the temporarily negatively charged carbon and adjacent carbocation form another bond to create the neutral propene product. Our data show that in the region of temperature that we investigated, the Gibbs free energy is always larger for 2-propanol compared to 1-propanol. The mechanism of removing H+ and grabbing it onto the water and the mechanism of creation of the double bond to form the propene requires more energy for the 2-propanol than for the 1-propanol. This leads the elimination mechanism E2 to be more in favor of 1-propanol reaction compared to 2-propanol. The following figures show the plots of the combined and separated different basis sets, cc-pVDZ, cc-pVTZ and cc-pVQZ at 398 K for the dehydration of 1- and 2- propanol.

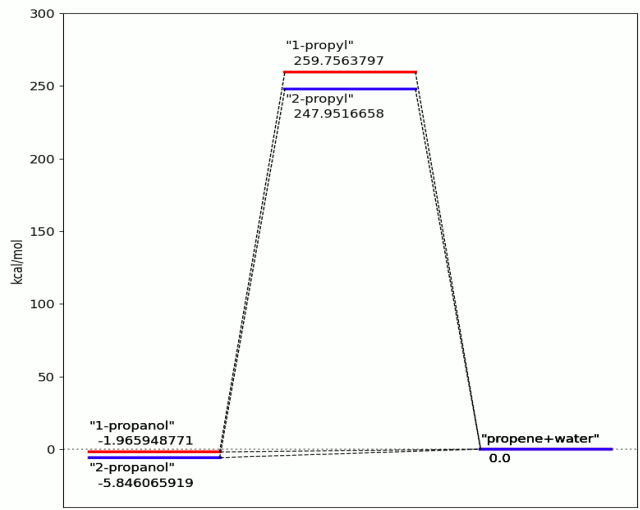

**Fig. 8-1. Dehydration of 1- and 2- propanol: combined different basis sets, cc-pVDZ, cc-pVTZ and cc-pVQZ all at 398 K.**

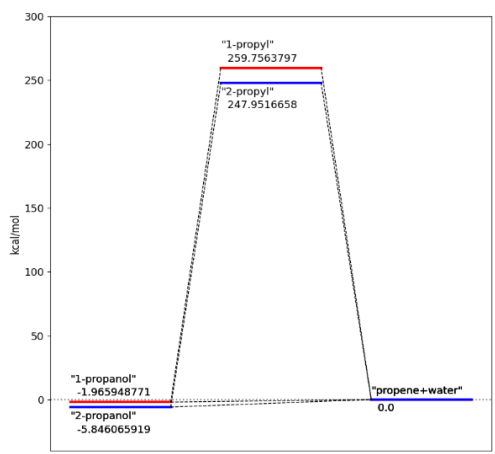

a)   cc-pVDZ

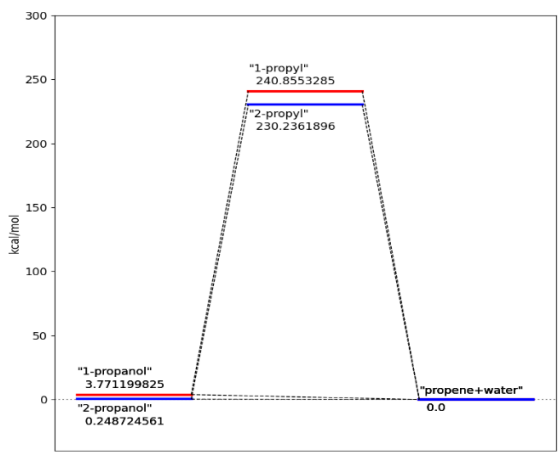

b)   cc-pVTZ

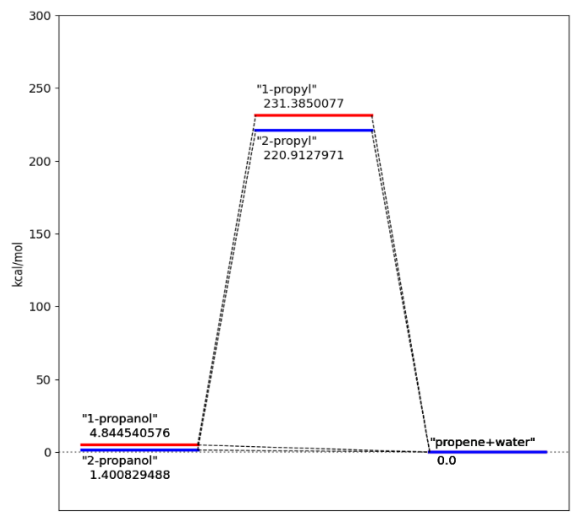

c)   cc-pVQZ

**Fig. 8-2. Dehydration of 1- and 2- propanol: separated different basis sets, cc-pVDZ, cc-pVTZ and cc-pVQZ all at 398 K.**

**Analysis and discussion**

      Figs. 8 show the Gibbs free energy results for E1 elimination mechanism during the dehydration of 1- and 2- propanol at different basis sets, cc-pVDZ, cc-pVQZ, and cc-pVTZ at 398 K. The E1 elimination is a first-order unimolecular reaction, meaning the rate depends only on the concentration of the molecule. As shown in Figs. 1-2 and 2-2, the E1 elimination reaction presents two step mechanisms that involve a carbocation intermediate. The first step is an ionization step, in which the hydroxide (OH) group breaks away from the molecule, leaving a positive carbocation intermediate. The leaving group takes a full octet of electrons with it, including the electron it was sharing with carbon to form the single bond. The second step is deprotonation of an adjacent carbon by a Lewis base. The base uses its valence electrons to grab onto the hydrogen. The C-H bond will release the hydrogen atom and the carbon will have extra electrons to share with the positively charged carbon next to it, creating a final neutral propene product. There were multiple hydrogens that the water could grab, but the most substituted propene product is favored. Fig. 8-1 presents the results of the combined dehydration of 1- and 2- propanol with three different basis sets, cc-pVDZ, cc-pVQZ, and cc-pVTZ, all at 398 K. Note that Fig. 8-1 is a moving a dynamics figure showing the behavior of the Gibbs free energy with 1- and 2- propanol in respect to the above three basis sets. We have set the Gibbs free energy to be zero for all products obtained. This allows to better investigate the Gibbs free energy of the reactants 1- and 2- propanol. The Gibbs free energy is negative for Fig. 8-2 a) for the calculations performed with cc-pVDZ basis set at 398 K. Under these computational conditions, the 1- and 2- propanol reactions are favorable and spontaneous because these reactants have more free energy than the products or final state, which is set in the graphs to zero. The 2-propanol reaction has more negative Gibbs free energy ($\Delta G = -5.846065919 \, KCal/mol$) compared to 1-propanol reaction ($\Delta G = -1.965948771 \, KCal/mol$), which means that under these computational conditions, the 2- propanol dehydration reaction is more favorable to happen or spontaneous than the 1-propanol reaction. The free energy available in the 1- and 2- propanol reaction for the E1 mechanism will be used during the formation of the intermediates 1- and 2- propyl. Furthermore, the available free energies of the two intermediates propyl are positive. Then, the energy would have to be added to the intermediate systems for $\Delta G$ to be smaller and drive the reaction to the products or final state. The 2- propyl free energy ($\Delta G = 247.9516658 \, KCal/mol$) is smaller than 1- propyl free energy ($\Delta G = 259.7563797 \, KCal/mol$), which means that the 2- propyl will require less energy than 1- propyl to make the reaction possible to the products or final state. The more negative Gibbs free energy of 2- propanol and the lower free energy of its intermediate 2- propyl compared to 1- propanol and its intermediate 1- propyl makes the dehydration of 2-propanol reaction more favorable and spontaneous compared to the dehydration of 1- propanol of the E1 elimination mechanism.

      Figs. 8-2 b) and c) represent the E1 elimination mechanism where an intermediate "propyl" is formed before the final products which are water and propene. The Gibbs free energies of reactants and intermediates are all positive. Under the above computational conditions, the dehydration of 1- and 2- propanol for E1 mechanism is endergonic, non-spontaneous, and not favorable. Because $\Delta G$ is less positive for 2-propanol reactant ($\Delta G = 1.400829488 \, KCal/mol$) and 2- propyl intermediate ($\Delta G = 220.9121 \, KCal/mol$) compared to 1-propanol reactant ($\Delta G = 4.844540576 \, KCal/mol$) and 1- propyl intermediate ($\Delta G = 231.3850077 \, KCal/mol$),

respectively, 2- propanol dehydration will require less energy compared to 1- propanol dehydration to make the process favorable. Comparing Figs. 8-2 b) and c), under the same thermodynamics' conditions, the calculations performed with the cc-pVQZ basis set of the dehydration of 1- and 2- propanol give lower Gibbs free energies compared to the calculations performed with the cc-pVTZ basis set. The following figures show the plots of the combined and separated different temperatures with increment of 100 K with the same basis sets, cc-pVQZ, for the dehydration of 1- and 2- propanol.

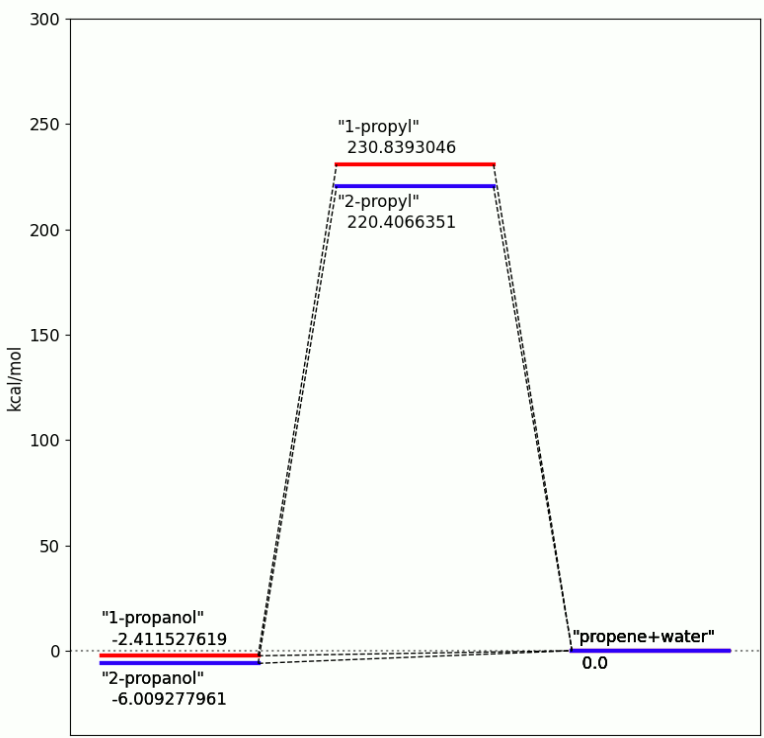

**Fig. 9-1. Dehydration of 1- and 2- propanol: combined images for cc-pVQZ basis set with increasing temperature (increment of 100K).**

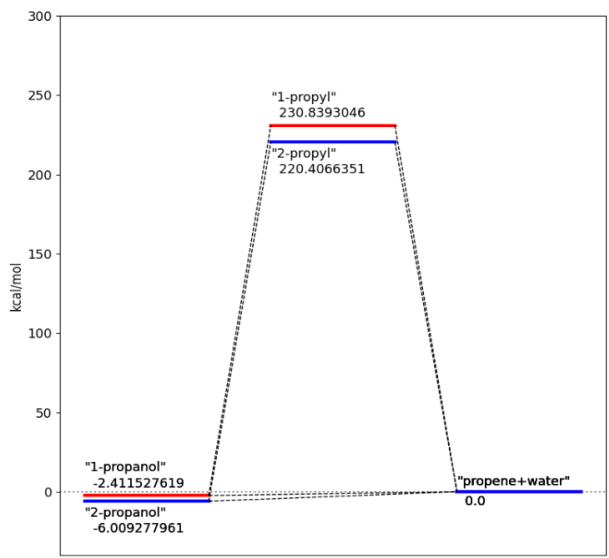 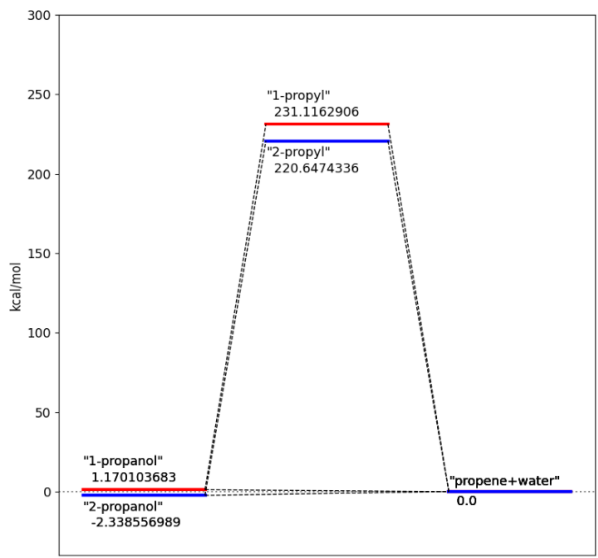

**a)     Images of cc-pVQZ basis set with 198 K.**    **b) Images of cc-pVQZ basis set with 298 K.**

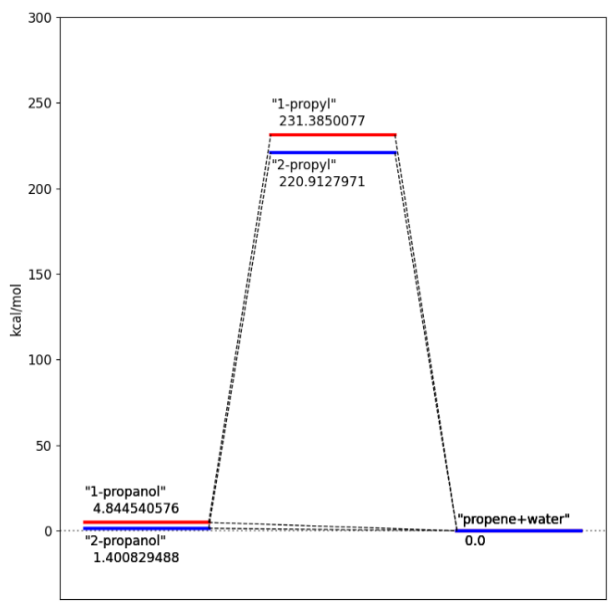
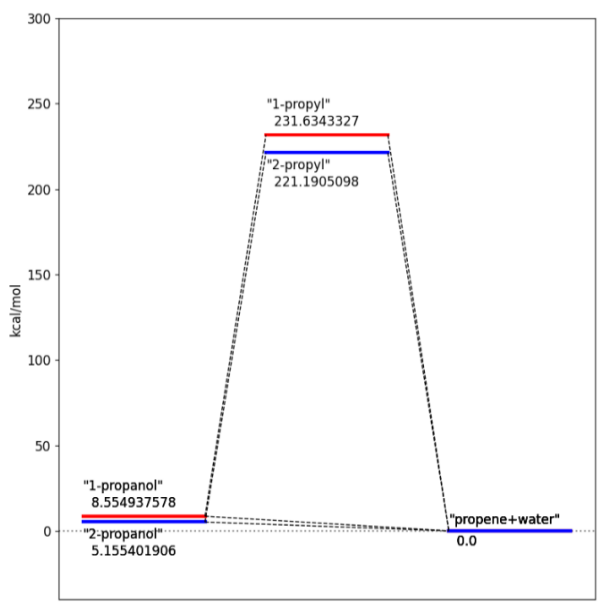

**c)     Images of cc-pVQZ basis set with 398 K.**    **d) Images of cc-pVQZ basis set with 498 K.**

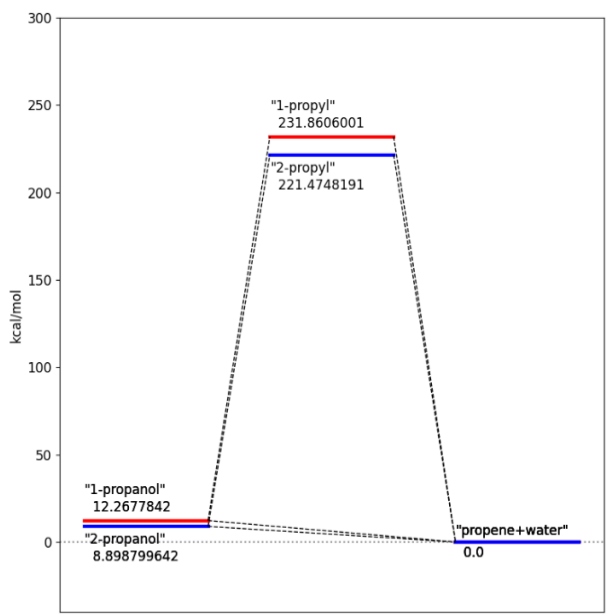
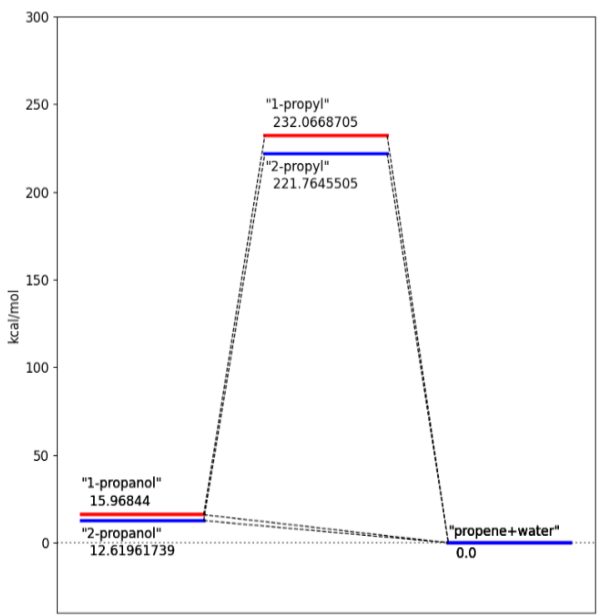

**e) Images of cc-pVQZ basis set with 598 K.**    **f) Images of cc-pVQZ basis set with 698 K.**

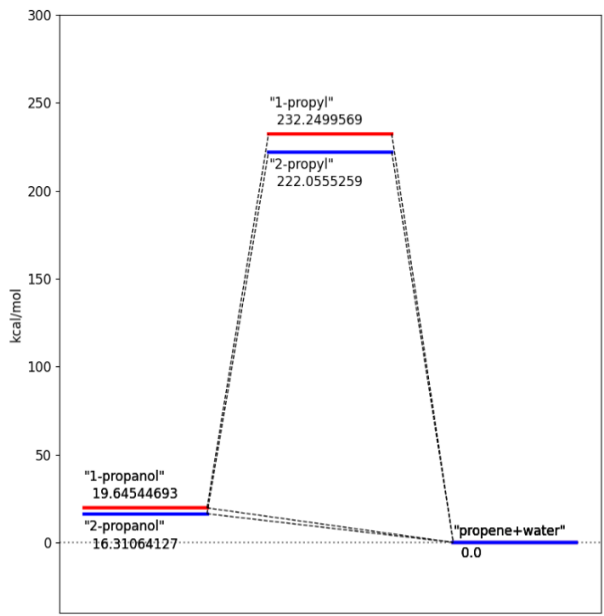
g) Images of cc-pVQZ basis set with 798 K.

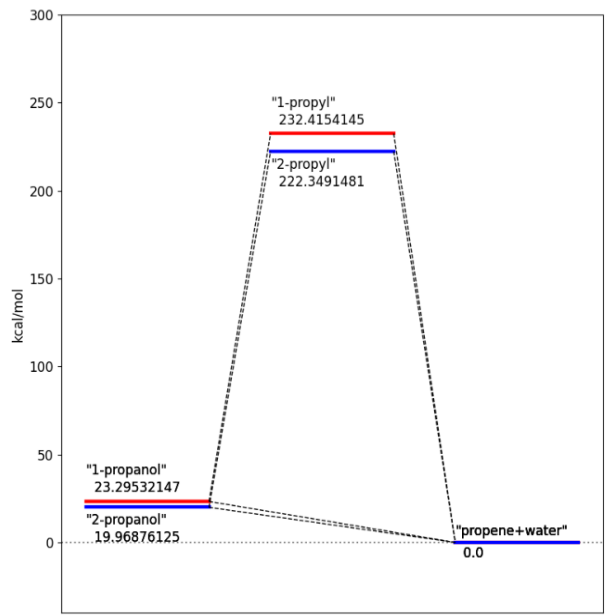
h) Images of cc-pVQZ basis set with 898 K.

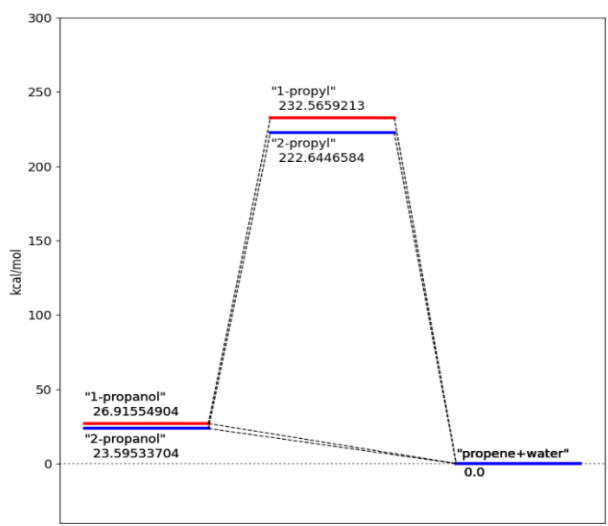
i) Images of cc-pVQZ basis set with 998 K.

**Figure 9-2. Dehydration of 1- and 2- propanol: separated images for cc-pVQZ basis set and increasing temperature: the temperature increases in steps of 100K from 198K to 998K.**

### Analysis and discussion

Figs. 9 show the Gibbs free energy results for the E1 elimination mechanism during the dehydration of 1- and 2- propanol for the favorable selected cc-pVQZ basis set with various temperatures ranging from 198 K to 998 K. As E1 elimination mechanism, all Figs. 9 (9-1 and 9-2) present two step mechanisms that include a carbocation intermediate. The plots of Figs. 9 show that the dehydration of 1- and 2- propanol reactions are favorable and spontaneous for E1 mechanism only at low temperatures. At 198 K (Fig. 9-2 a)), the Gibbs free energies are negative ($\Delta G < 0$) for both 1- and 2- propanol reactions, with the 2- propanol reactant having the more negative free energy ($\Delta G = -6.009277961\ KCal/mol$) compared to 1- propanol reactant ($\Delta G = -2.411527619\ KCal/mol$). As such, the 2- propanol reaction is more favorable and spontaneous than 1- propanol reaction at 198 K, and this is supported by the fact that the intermediate 2- propyl ($\Delta G = 220.4066351\ KCal/mol$) derived from 2- propanol reactant will require less energy than 1- propyl ($\Delta G = 230.8393046\ KCal/mol$) to make the process favorable and spontaneous. Furthermore, at 298 K (Fig. 9-2 b)), the dehydration 2- propanol reaction is still favorable and spontaneous ($\Delta G = -2.338556989\frac{KCal}{mol} < 0$) while the 1- propanol reactant ($\Delta G = 1.170103683\frac{KCal}{mol} > 0$) will require energy to make the reaction favorable and spontaneous. As temperature increases with an increment of 100 K from 398 K (Fig. 9-2 c)) to 998 K (Fig. 9-2 i)), both reactions 1- and 2- propanol become more and more non-favorable and non-spontaneous. The Gibbs free energies ($\Delta G < 0$) is positive for temperatures above 298 K in the regime of temperature investigated in this work. The dehydration of 2- propanol reactant in the regime of $\Delta G > 0$ requires less energy than the dehydration of 1- propanol to make the reaction favorable and spontaneous for E1 elimination mechanism.

## 1.5. Concluding Remarks

The NWChem project is an example of a successful co-design effort that harnesses the expertise and experience of researchers in several complementary areas, including quantum chemistry, applied mathematics, and high-performance computing. Over the last three decades, NWChem has evolved into a code that offers a unique combination of computational tools to tackle complex chemical processes at various spatial and time scales. Besides developing new methodologies, NWChem is consistently enhanced with innovative algorithms to leverage emerging computer architectures and quantum technologies. We believe that the community model of NWChem will continue to spur exciting new developments well into the future. In this work, we established how the accuracy of the calculations depends on the basis sets, and we determined what basis sets are needed to achieve sufficient accurate results. We also calculated the reaction free energy as a function of temperature as thermodynamic parameter. We found that at low temperature the reaction is thermodynamically unfavorable. Therefore, high temperatures are needed to drive this reaction. We have begun exploring the reaction mechanism at the zeolite's active site and the role of detailed electronic structure calculations in enhancing our understanding and potential improvement of catalysts. Future publications will detail these research efforts. NWChemEx contains an umbrella of modules that can be used to tackle most electronic structure theory calculations being carried out today. We are working on setting up measures for chemical systems embedded in a larger system. This work opens ways to understand how DFT can be used for the removal of water from organic compounds (2-propanol) to form new chemicals (over H-ZSM-5 zeolite). Our goals are to lay the groundwork for the mechanistic simulations and study the energy at different stages and transition states. Further insights into catalytic properties could be obtained in the future by comparing DFT calculations to other corresponding methods of predictions. Our results show how the enthalpy of each basis set decreases as the number of valence orbitals increases. The findings in this article will form the basis for comparing calculated structures with experimental data. This work will explore and inform the NWChemEx development roadmap.

## 1.6. Open Questions and Challenges

Many open questions remain, such as how does the zeolite structure affect chemistry or how to understand the dynamics and mechanism? The Comparison between computational and experimental results is also very important. Correlated advanced solid-state NMR and DFT analyses will yield new molecular-level insights on the chemical bonds of materials used in battery research[14]. We employed DFT to compute the NMR properties of the ZSM-5 zeolite catalyst. We also plan to extend DFT calculations for NMR properties of electrolyte materials to derive universal fundamental insights into the structure and composition of superionic materials. The goal is to create a groundbreaking research program focused on computing NMR properties of molecules and materials[10-13].

Future research should focus on enhancing the convergence of quantum chemistry methods, particularly by developing more robust basis sets that can accurately predict material properties. Achieving chemical accuracy involves minimizing the error in reaction enthalpy to less than 1 kcal/mol. For our zeolite system, we essentially need to optimize the initial full structure, generate truncated fragments, terminate the model structure, and then have the system react subject to a constraining environment. Next, we need to carve out a cluster of the zeolite around the active site. The active site is the AlOH piece, and we need to include some of the silicates around that. We need to terminate this cluster with hydrogens and optimize the positions of the hydrogens.

## 1.7. Perspectives: Towards an Improved Fundamental Understanding of Mechanistic Simulations of Catalysts

The goals of this axe of research are to,

1. Establish the mechanism of 2-propanol dehydration over ZSM-5 to produce propene and water.

    a. Water is a polar solvent which is likely to introduce significant solvent effects.
    b. Water is likely playing a role in this mechanism and this role needs to be established.
    c. Identifying reaction conditions, especially the temperatures at which reactions are thermodynamically favored, is crucial.

2. Design ab-initio molecular dynamics simulations for NWChemEx to investigate the effect of the confinement in the zeolite cages on the reaction.

NMR spectroscopy is an experimental technique that can be used to verify predictions from the ab-initio simulations.

    a. We plan to compare DFT simulations of finite clusters using NWChem's Gaussian basis set GIAO codes and Quantum Espresso's periodic GIPAW code. The key here is to understand the convergence of the calculations as a function of different types of basis sets, Gaussian basis functions vs. plane waves.

b. We plan to compute NMR spectra for periodic zeolite models to compare to experiment.

Improvements in the accuracy of density functionals are likely to enhance the reliability of computed NMR parameters across various material classes. Such advances are of urgency for the treatment of systems with strongly localized orbitals and methods that can incorporate nonlocal functionals might be beneficial. Such functionals are extremely expensive in plane-wave calculations. The NWChemEx package might provide an efficient route to computing shielding tensors with such functionals.

**Acknowledgements:** *This research was supported by the Exascale Computing Project (17-SC-20-SC), a collaborative effort of two U.S. Department of Energy organizations (Office of Science and the National Nuclear Security Administration) responsible for the planning and preparation of a capable exascale ecosystem, including software, applications, hardware, advanced system engineering and early testbed platforms, in support of the nation's exascale computing imperative.*